\documentclass[doublecol]{epl2} 

\usepackage{here}
\usepackage{amsmath}         
\usepackage{graphicx}
\usepackage{braket}         
\usepackage{color}
\usepackage[utf8]{inputenc}
\DeclareUnicodeCharacter{2212}{-}

\title{Orbitronics: Orbital Currents in Solids}
\shorttitle{Orbitronics: Orbital Currents in Solids} 

\author{Dongwook Go\inst{1,2}\thanks{E-mail: \email{d.go@fz-juelich.de}} \and Daegeun Jo\inst{3} \and Hyun-Woo Lee\inst{3} \and Mathias Kl\"aui\inst{2,4,5} \and Yuriy Mokrousov\inst{1,2}}
\shortauthor{D. Go \etal}

\institute{                    
  \inst{1} Peter Gr\"unberg Institut and Institute for Advanced Simulation, Forschungszentrum J\"ulich and JARA, 52425 J\"ulich, Germany \\
  \inst{2} Institute of Physics, Johannes Gutenberg University Mainz, 55099 Mainz, Germany \\
  \inst{3} Department of Physics, Pohang University of Science and Technology, Pohang 37673, Korea \\
  \inst{4} Graduate School of Excellence Materials Science in Mainz, 55128 Mainz, Germany \\
  \inst{5} Center for Quantum Spintronics, Department of Physics, Norwegian University of Science and Technology, NO-7491 Trondheim, Norway
}

\pacs{75.76.+j}{Spin transport effects}
\pacs{73.40.−c}{Electronic transport in interface structures}
\pacs{72.90.+y}{Other topics in electronic transport in condensed matter}

\abstract{
In solids,  electronic Bloch states are formed by atomic orbitals. While it is natural to expect that orbital composition and information about  Bloch states can be manipulated and transported, in analogy to the spin degree of freedom  extensively studied in past decades, it has been assumed that orbital quenching by the crystal field prevents significant dynamics of orbital degrees of freedom. However, recent studies reveal that an orbital current, given by the flow of electrons with a finite orbital angular momentum, can be electrically generated and transported in wide classes of materials despite the effect of orbital quenching in the ground state. Orbital currents also play a fundamental role in the mechanisms of other transport phenomena such as spin Hall effect and valley Hall effect.
Most importantly, it has been proposed that  orbital currents can be used to induce magnetization dynamics, which is one of the most pivotal and explored aspects of magnetism. Here, we give an overview of recent progress and current status of research on orbital currents. We review proposed physical mechanisms for generating orbital currents and discuss candidate materials where orbital currents are manifest. We review recent experiments on orbital current generation and transport, and discuss various experimental methods to quantify this elusive object at the heart of \emph{orbitronics} $-$ an area which exploits the orbital degree of freedom as an information carrier in solid-state devices.
}

\begin{document}
\maketitle

\section{Introduction}

Atomic orbitals and their linear combinations present elementary building blocks for electronic Bloch states in solids. 
Given a specific Bloch vector $\mathbf{k}$, the index $n$ of a Bloch state $\psi_{n\mathbf{k}}$ carries information about two key aspects: valence orbitals of the constituent atoms and intrinsic spin of the electron. Consequently, these two aspects of Bloch states promote  degrees of freedom which can be manipulated and transported for novel types of solid-state device applications. In fact, studies of the spin degree of freedom inherent to Bloch states have evolved into a prominent area of spintronics, which addresses  generation, detection, and manipulation of the spin information \cite{Zutic2004, Fert2008, Sinova2015, Manchon2019}. Similarly, one may think of the orbital degree of freedom as a variable which can be controlled in solids, referring to the respective branch of condensed matter physics as \emph{orbitronics} \cite{Bernevig2005, Go2017, Phong2019}.
So far, however,  orbital transport has been investigated to a much lesser degree than the spin transport. It is because the orbital degree is often regarded as ``frozen'' in solids: since the crystal field enforces specific symmetry of the Bloch states, this breaks the continuous rotation symmetry and suppresses the formation of the orbital moment \cite{Kittel_textbook}. Thus, for elemental $3d$ ferromagnets (FMs), the orbital moment is much smaller than the spin moment as it requires spin-orbit coupling (SOC), which is much weaker than the crystal field potential \cite{Meyer1961}. However,  exceptions exist, and it is mainly the  orbital moment which contributes to  magnetism in some materials~\cite{Shindou2001, Hernando2006, Hanke2017, Tschirhart2021}. 

In contrast, recent studies have shown that orbital quenching does not preclude prominent nonequilibrium dynamics and transport of the orbital degree of freedom \cite{Go2018, Go2021b}, which is a property of the excited state rather than of the ground state. While crystal field potential has the strongest influence on the properties of electronic states in solids, it also mediates a hybridization between different atomic orbitals, which is a crucial element in  orbital dynamics \cite{Go2018}. This implies that the dynamics and transport of orbital information can be driven by external stimuli, such as electric field, regardless of the orbital quenching governing the ground state. Recent studies have indeed demonstrated not only that pure orbital transport, which does not involve other (e.g. spin) currents, takes place in many materials \cite{Bernevig2005, Go2018, Jo2018, Phong2019, Bhowal2020a, Bhowal2020b, Canonico2020a, Canonico2020b, Cysne2021, Sahu2021, Mu2021}, but also that orbital currents mediate other transport phenomena such as spin Hall effect (SHE) \cite{Tanaka2008, Kontani2009, Go2018} and valley Hall effect (VHE) \cite{Cysne2021, Bhowal2021}. Moreover, recent theories suggest that an injection of the orbital current into a FM can excite magnetization dynamics \cite{Go2020a, Go2020b, Go2021b}, which is one of the most important functions required for spintronic devices. This possibility has been explored in recent experiments, which demonstrated an alternative way of using the orbital current instead of the spin current in spintronics \cite{Ding2020, Kim2021, Lee2021a, Lee2021b, Ding2021, Tazaki2020}.

%

In this Perspective, we introduce the notion of the orbital current in solids and key orbital transport phenomena, which enable an electrical generation and detection of  orbital currents. We also explain and develop the idea of utilizing  orbital currents and orbital injection as  means to manipulate local moments in magnetic materials. 
Given that the experimental realization and observation of orbital effects has only recently started, we put a strong emphasis on reviewing experimental methods for the detection of the orbital currents. Finally, we discuss challenges and future directions in orbitronics which utilize orbital currents as an information carrier in solid-state devices and resulting effects.

\section{Orbital current}

Theoretical description of the orbital current is similar to that of the spin current. However, a clear difference is that an orbital current cannot be defined in the vacuum unlike the spin current which is often schematically depicted as a flow of  ``arrows''. The notion of an orbital current is based on the orbital character of the Bloch states in a solid, which comprises many atoms. Thus, it is an \emph{emergent} concept which does not exist for  constituent elements per se \cite{Anderson1972}. In the wave packet description of $p$-orbital states, for example, an arbitrary quantum state can be written as 
\begin{eqnarray}
\ket{\Psi (t)}
&=&
\int d\mathbf{k}
\left[
C_{p_x\mathbf{k}}(t)
\ket{\psi_{p_x\mathbf{k}}}
\right.
\\
& &
\left.
+
C_{p_y\mathbf{k}}(t)
\ket{\psi_{p_y\mathbf{k}}}
+
C_{p_z\mathbf{k}}(t)
\ket{\psi_{p_z\mathbf{k}}}
\right],
\nonumber
\end{eqnarray}
where $\ket{\psi_{p_\alpha \mathbf{k}}}$ is the Bloch state derived from the $p_\alpha\ (\alpha=x,y,z)$ orbital of the atomic state. Here,  orbital information is encoded in coefficients $C_{p_\alpha\mathbf{k}}(t)$, which can be controlled by external perturbations. For instance, a wave packet with $\mathrm{arg}(C_{p_y\mathbf{k}}/C_{p_x\mathbf{k}})=\pi/2$ and $C_{p_z\mathbf{k}}=0$ carries a finite orbital angular momentum (OAM) along $z$ direction. Note that the coefficients behave as an ``internal'' degree of freedom of the wave packet, analogously to the case of spin. Thus, we can define the OAM operator in a matrix representation, and the orbital current operator can be written as a second rank tensor
\begin{eqnarray}
j^{L_\beta}_\alpha 
=
\frac{1}{2}
(v_\alpha L_\beta + L_\beta v_\alpha),
\end{eqnarray}
where $v_\alpha$ is $\alpha$-component of the velocity operator and $L_\beta$ is $\beta$-component of the OAM operator. Note that $\mathbf{L}$ refers to the intra-atomic contribution to the OAM, which does not contain the inter-atomic contribution. Except for a particular case \cite{Bhowal2021}, a general definition of the orbital current that incorporates the intra- and inter-atomic contributions on an equal footing is not known yet. Theories for transport of OAM which bridge the two pictures (localized and delocalized) need to be developed, which might require a further effort from the side of the  Berry phase and general response theories \cite{Thonhauser2005, Shi2007, Essin2009, Malashevich2010}.

\section{Orbital Hall effect}

In analogy to the definition of the SHE \cite{Dyakonov1971, Hirsch1999, Sinova2015}, the orbital Hall effect (OHE) refers to a generation of a transverse orbital current by an external electric field \cite{Bernevig2005, Kontani2009, Go2018}. The OHE serves as a way to electrically generate an orbital current. The reciprocal effect $-$ the inverse OHE $-$ can be used to detect orbital currents electrically. 

\subsection{Mechanisms}

Like many other Hall effects, mechanisms behind the OHE can be divided into \emph{intrinsic} and \emph{extrinsic} types. The intrinsic mechanism, which originates from the ground state electronic states, has been proposed for semiconductors \cite{Bernevig2005}, metals \cite{Tanaka2008, Kontani2009, Go2018, Jo2018}, and 2D materials \cite{Tokatly2010, Bhowal2020a, Bhowal2020b, Canonico2020a, Canonico2020b, Cysne2021, Bhowal2021}. However, the picture of an extrinsic mechanism, which relies on disorder scattering, has not been developed to date, although it is expected to be present, e.g. in the form of side jump and skew-scattering mechanisms. Only in very specifics situations, a consequence of the vertex corrections has been considered. Bernevig \emph{et al.} showed that vertex corrections from  impurity scattering vanish for a low-energy model for hole-doped Si \cite{Bernevig2005}. Vertex corrections in transition metals are in general finite, but Tanaka \emph{et al.} showed that they are expected to be small \cite{Tanaka2008}.
Here we review only the intrinsic mechanism in detail. 

\begin{figure}[t!]
\centering
\includegraphics[angle=0, width=0.49\textwidth]{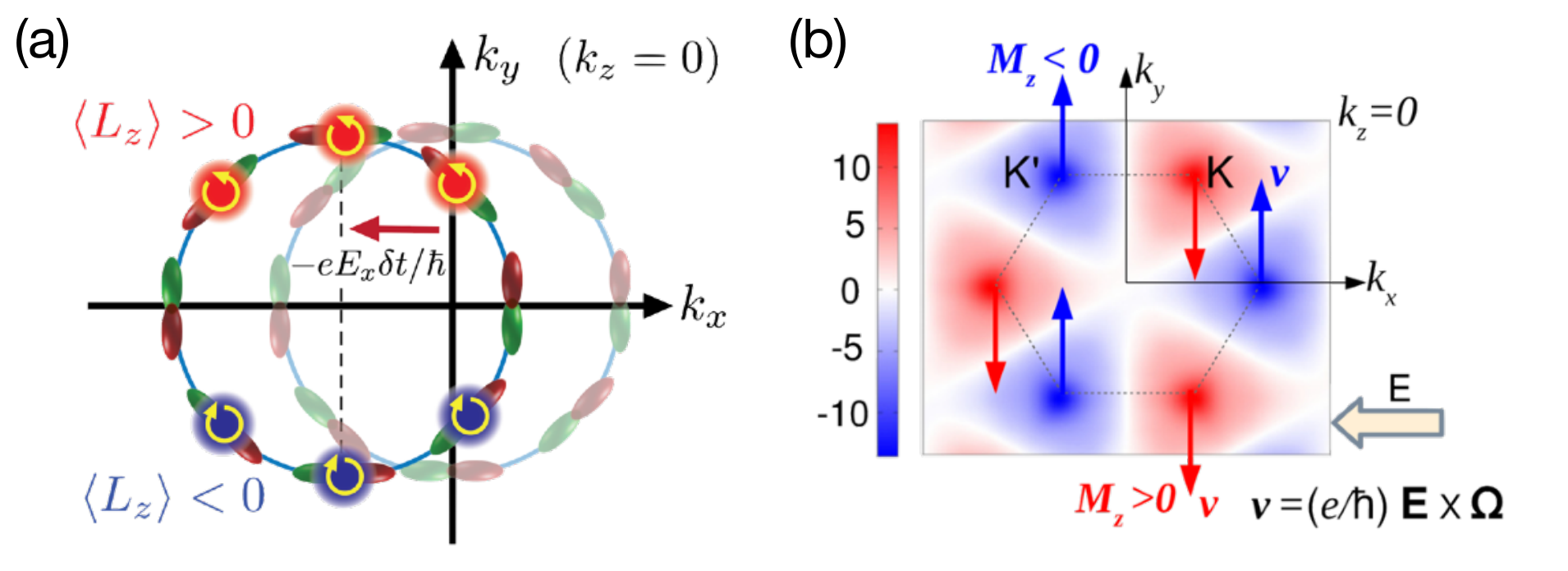}
\caption{
\label{fig:OHE_mechanisms}
Mechanisms of the OHE. (a) In centrosymmetric crystals,  equilibrium OAM is vanishing at each $\mathbf{k}$, but an external electric field can induce finite OAM along the direction $\mathbf{k}\times\mathbf{E}$. Such $\mathbf{k}$-dependent induced OAM generates the OHE. (b) In a TMD monolayer, broken spatial inversion symmetry gives rise to a valley-dependent OAM in equilibrium (indicated by color). Thus, the anomalous velocity (indicated by arrows) driven by the Berry curvature, which is also valley-dependent, generates the Hall current which is already orbitally-polarized. Adapted from Refs. \cite{Go2018, Bhowal2020a}.
}
\end{figure}

The intrinsic contribution depends on whether the spatial inversion symmetry is present in the system or not. Time-reversal and spatial inversion symmetries imply $\langle \mathbf{L} \rangle_\mathbf{k} = -\langle \mathbf{L} \rangle_{-\mathbf{k}}$ and $\langle \mathbf{L} \rangle_\mathbf{k} = \langle \mathbf{L} \rangle_{-\mathbf{k}}$, respectively. In nonmagnetic materials, a combination of  time-reversal symmetry and spatial inversion symmetry dictates that OAM is absent for all states at each $\mathbf{k}$ in the ground state, which is a ``strong'' condition of orbital quenching. Thus, in centrosymmetric crystals,  OAM has to be induced \emph{a priori} by an external electric field. Go \emph{et al}. have demonstrated that an external electric field can drive a hybridization between states with different orbital character, e.g. radial or tangential orbitals, which can induce a finite OAM \cite{Go2018}. As shown in Fig.~\ref{fig:OHE_mechanisms}(a), an important feature of the orbital hybridization mechanism is that an OAM is induced along the direction of $\mathbf{k}\times \mathbf{E}$ in the lowest order in $\mathbf{k}$, where $\mathbf{E}$ is an external electric field. This means that electrons with opposite sign of  induced OAM propagate in opposite directions, which is effectively the OHE.  

On the other hand, in noncentrosymetric crystals, OAM may already be present in equilibrium at each $\mathbf{k}$-point although the total OAM vanishes in the presence of  time-reversal symmetry (``weak'' condition of orbital quenching). At the same time, it is important to note that the Berry curvature of Bloch states, which gives rise to the anomalous velocity, satisfies the same symmetry constraints as OAM,~i.e.~the direction of the Berry curvature field is correlated with the direction of the OAM. Thus, the Hall current driven by the Berry curvature is naturally orbitally-polarized. This mechanism was pointed out by Bhowal and Satpathy \cite{Bhowal2020a, Bhowal2020b}. In transition metal dichalcogenide (TMD) monolayers, orbital polarization is mostly pronounced at $\mathrm{K}$ and $\mathrm{K}'$ valleys, where the sign of OAM is opposite (Fig.~\ref{fig:OHE_mechanisms}(b)). In addition, the Berry curvature is also valley-dependent, as it is well-known from the studies of valley Hall effect (VHE) \cite{Xiao2007, Schaibley2016, Vitale2018, Mak2018}. Therefore, TMD monolayers can exhibit the OHE, where the mechanism is distinct from that active for centrosymmetric materials. However, we emphasize that 
both mechanisms can co-exist in noncentrosymmetric materials in general \cite{Sahu2021}.

\subsection{Relation to other Hall effects}

\begin{figure}[t!]
\centering
\includegraphics[angle=0, width=0.47\textwidth]{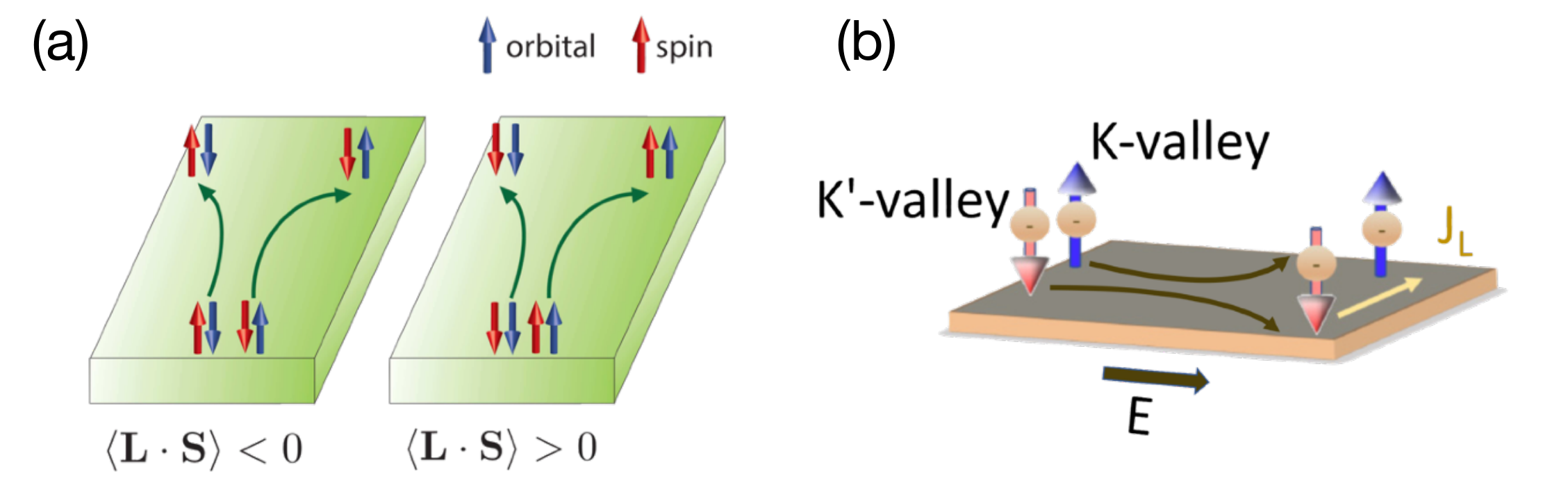}
\caption{
\label{fig:OHE_SHE_VHE}
(a) Relative sign between the OHE and the SHE is determined by the correlation $\langle \mathbf{L}\cdot\mathbf{S} \rangle$ mediated by spin-orbit coupling. (b) In 2D materials without inversion symmetry, such as a TMD monolayer, the VHE is associated with the OHE. Adapted from Refs. \cite{Go2018, Bhowal2021}.
}
\end{figure}

The OHE can play a crucial role in other Hall effects. One of the most important examples of this is the fact that the SHE results from the OHE. We emphasize that the mechanism of the OHE does not require SOC in general \cite{Bernevig2005, Kontani2009, Go2018}. This means that the OHE can be finite in materials with negligible SHE \cite{Jo2018}. When SOC is taken into account, OAM and spin become entangled and OHE is accompanied by SHE \cite{Tanaka2008, Go2018}. The relative sign between the OHE and SHE generally depends on the correlation $\langle \mathbf{L}\cdot\mathbf{S}\rangle$, where $\mathbf{S}$ is the operator of spin, as shown in Fig.~\ref{fig:OHE_SHE_VHE}(a) \cite{Go2018}. This explains the overall trend in sign and magnitude of  SHE in $4d$ and $5d$ transition metals \cite{Tanaka2008, Kontani2009}. Since the SHE originates from the OHE, i.e. the SHE vanishes in the absence of the OHE, the strength of SHE is fundamentally limited by the strength of the OHE. Thus, finding large-SHE materials, which is one of important challenges in spintronics, is closely tied with finding materials which exhibit a large OHE. Meanwhile, Kontani \emph{et al.} demonstrated that the OHE plays a critical role for the anomalous Hall effect as it originates from the SHE when occupations of majority and minority electrons differ due to the exchange splitting \cite{Nagaosa2010, Sinova2015}. 

In 2D materials with honeycomb lattice, the VHE arises due to sublattice symmetry breaking, which is achieved e.g. in TMD monolayers or graphene grown on hexagonal BN. As discussed in Fig.~\ref{fig:OHE_mechanisms}(b), the valley-dependent Berry curvature is entangled with the OAM, and thus the VHE is tied to the OHE. However, the valley information is not a physically measurable quantity and its definition is ambiguous, depending on an arbitrary cutoff of the $\mathbf{k}$-space integral. For this reason, Bhowal and Vignale proposed that the VHE is better described by the OHE \cite{Bhowal2021}. However, for disentangling the OHE and the VHE, Cysne \emph{et al.} proposed to use TMD bilayers where the inversion symmetry is present, which eliminates the valley-dependent Berry curvature and eventually the VHE while keeping the OHE finite \cite{Cysne2021}. 
\subsection{Materials}

\begin{figure}[t!]
\centering
\includegraphics[angle=0, width=0.49\textwidth]{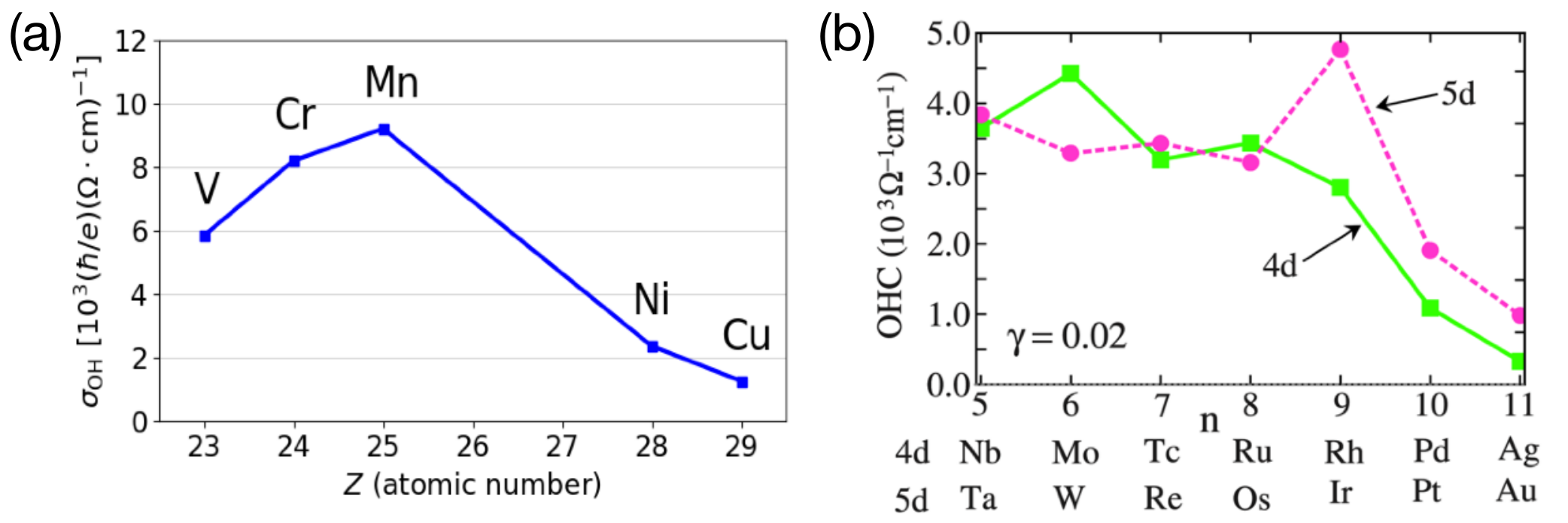}
\caption{
\label{fig:OHE_TMs}
The OHCs of transition metals for (a) $3d$, and (b) $4d$, $5d$ elements. We remark that the magnitude of the OHC is generally an order of magnitude larger than the SHC for $4d$ and $5d$ elements (not shown). Adapted from Refs. \cite{Jo2018, Tanaka2008}.
}
\end{figure}

So far, classes of materials for which the OHE have been investigated are very limited. Bernevig \emph{et al.} theoretically found that OHE is sizeable in hole-doped Si despite negligible SOC, which originates in the $p$ orbital character of the valence bands \cite{Bernevig2005}. For transition metals, Jo \emph{et al.} studied $3d$ elements and found that the orbital Hall conductivities (OHCs) are gigantic although the spin Hall conductivities (SHCs) are much smaller due to small SOC \cite{Jo2018} (Fig.~\ref{fig:OHE_TMs}(a)). Tanaka and Kontani \emph{et al.} investigated $4d$ and $5d$ transition metals and found that the OHCs are also huge (Fig.~\ref{fig:OHE_TMs}(b)), which leads to sizeable SHCs \cite{Tanaka2008, Kontani2009}.

\begin{figure}[b!]
\centering
\includegraphics[angle=0, width=0.49\textwidth]{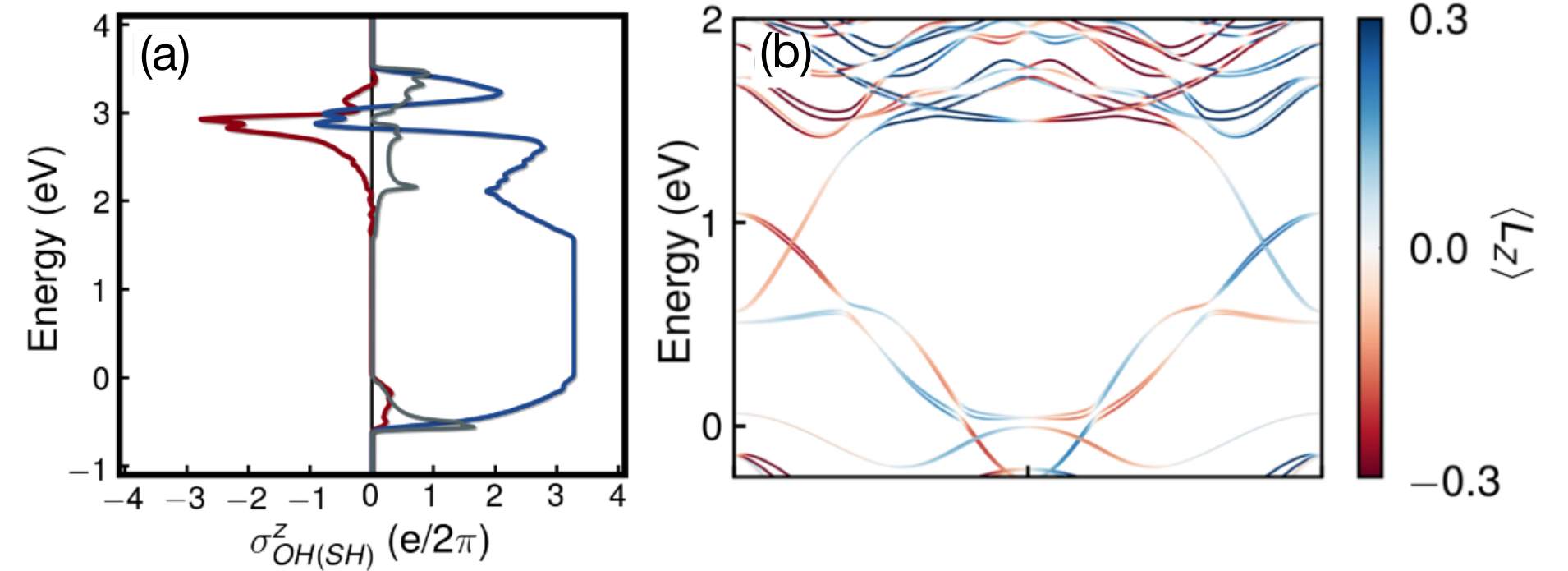}
\caption{
\label{fig:OHE_insulator}
(a) The OHC and SHC in a monolayer MoS$_2$, which are indicated by blue and red curves, respectively. The grey curve indicates the density of states. (b) Electronic states in a nanoribbon geometry with the zigzag boundary, where the edge states are orbitally-polarized (color). The energy reference is defined with respect to the valence band maximum. Adapted from Refs. \cite{Canonico2020b, Cysne2021}.
}
\end{figure}

The OHE in 2D materials such as TMDs has been also investigated by Canonico \emph{et al.} \cite{Canonico2020a, Canonico2020b} independently from Refs. \cite{Bhowal2020a, Bhowal2020b}.  In particular, it was found that OHC is finite even within the energy gap of TMDs (Fig.~\ref{fig:OHE_insulator}(a)) \cite{Canonico2020b}. This is particularly intriguing since the $Z_2$ topological invariant was found to be trivial in the latter case, as reflected in a zero SHC within the energy gap. Soon after, Cysne \emph{et al.} showed that the orbital Hall insulating phase is characterized by a nonzero orbital Chern number \cite{Cysne2021} by generalizing the method developed for computing the spin Chern number \cite{Prodan2009}. They also calculated electronic states in the TMD nanoribbon geometry with a zigzag boundary. The result seems to indicate that a nonzero orbital Chern number in the bulk corresponds to existence of the edge states which are orbital-polarized (Fig.~\ref{fig:OHE_insulator}(b)) \cite{Cysne2021}. However, it is still an open question whether an orbital Hall insulator is just another manifestation of an already-known topological phase or whether it goes beyond the current paradigm of topological insulators, and its physical meaning and consequences for  orbital transport need to be investigated further in depth. We also note that Tokatly investigated the OHE in hole-doped graphene and found that it is related to the Berry phase of the degeneracy point at $\mathbf{k}=0$ \cite{Tokatly2010}.  

\subsection{Experimental detection}

A direct way to experimentally confirm the OHE is by measuring the OAM accumulation at the boundary of the sample. For example, this can be done by the magneto-optical Kerr effect (MOKE), which has been previously employed to measure the spin accumulation caused by the SHE \cite{Kato2004, Stamm2019}. Because the photon dominantly interacts with the orbital part of the wave function and the spin is indirectly measured by the presence of the SOC, performing similar types of experiments on light elements with negligible SOC can be a way to detect the OAM accumulation.

However, in principle, the spin accumulation is always accompanied by the OAM accumulation, and it is not easy to distinguish them using MOKE. The X-ray magnetic circular dichroism (XMCD) provides a way to separately quantify the spin and orbital contributions \cite{vanderLaan2014}. However, since the X-rays can easily penetrate through thin films in a transmission geometry and there is no net induced OAM in a stand-alone sample, it is necessary to deposit an additional layer. Stamm \emph{et al.} attempted to measure the transient OAM of a nonmagnet (NM) in Pt/NM bilayers, where Ti, Cr, Cu were chosen as NMs \cite{Stamm2019}. However, no appreciable signal was observed within the detection limit of the experimental setup, from which the upper limit of the OAM accumulation was estimated to be $3\times 10^{-6}\mu_\mathrm{B}$ per atom. This value might appear small, but we stress that a theoretical calculation predicts that the induced OAM is an order of magnitude larger than the induced spin \cite{Salemi2020}.

We emphasize that just as the polarized light is used for obtaining MOKE or XMCD signal, any probe that interacts with the magnetic moment can be used to detect the OHE. For example, vortex beams \cite{Allen1992}, which carry finite OAM in addition to the spin, are expected to efficiently interact with the OAM in solids. Although vortex beams have been demonstrated in various areas of physics \cite{Shen2019}, it has been recently applied to the spectroscopy of magnetic materials \cite{Sirenko2019}. Using the vortex beam to detect transient angular momentum in solids needs to be explored further.

\begin{figure}[b!]
\centering
\includegraphics[angle=0, width=0.49\textwidth]{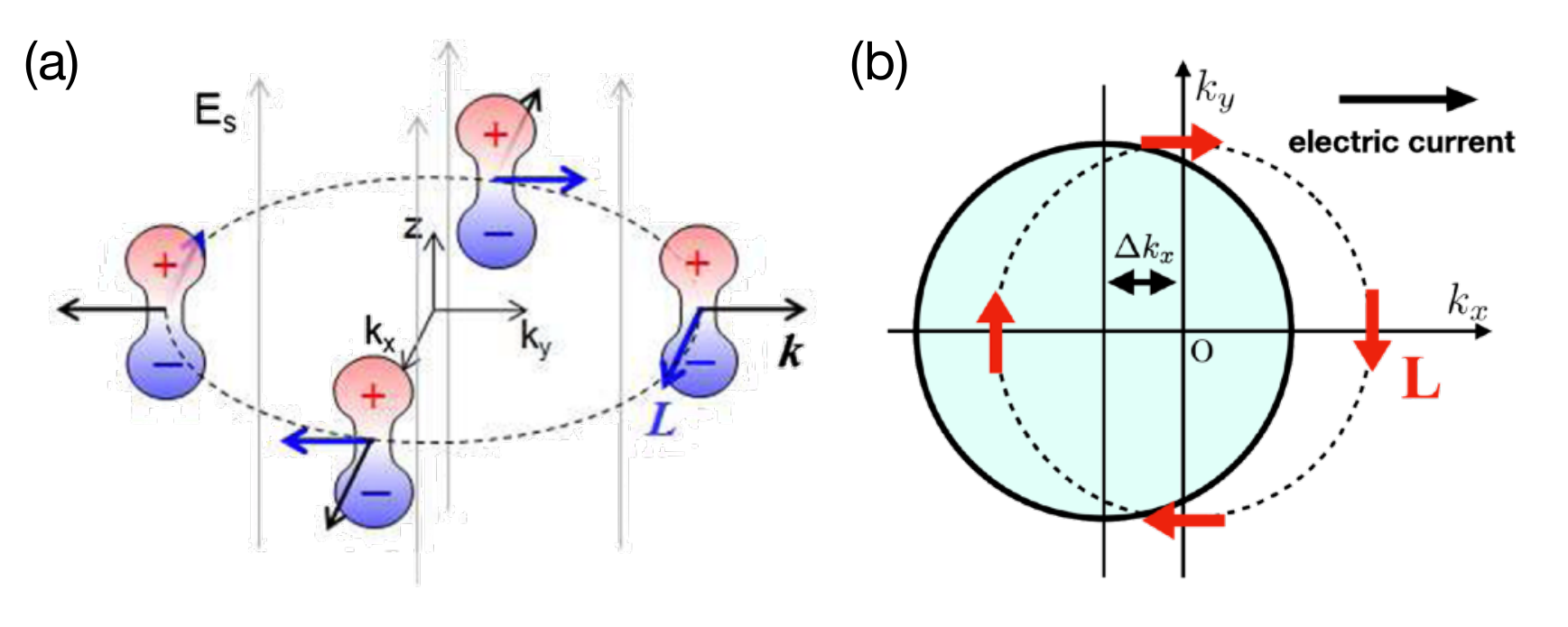}
\caption{
\label{fig:ORE} (a) A mechanism of the ORE. An electric dipole moment $\sim\mathbf{L}\times\mathbf{k}$ couples to a surface electric field $\mathbf{E}_s$. (b) OEE in the presence of a chiral OAM texture in $\mathbf{k}$-space. Adapted from Refs. \cite{Park2011, Kim2021}.
}
\end{figure}

\section{Orbital Rashba effect}

While the OHE is a mechanism prominent in the bulk of materials, the orbital Rashba effect (ORE), found at surfaces and interfaces, can also result in current-induced OAM accumulation \cite{Go2017, Yoda2018, Salemi2019, Johansson2021}. The ORE denotes a Rashba-type coupling between $\mathbf{L}$ and $\mathbf{k}$, which leads to orbital-dependent energy splitting and chiral OAM textures in $\mathbf{k}$-space \cite{Park2011, Park2013, Go2017}. The ORE emerges from the formation of electric dipole moment $\sim \mathbf{L}\times\mathbf{k}$, which couples to a potential gradient present at a surface or at an interface (Fig.~\ref{fig:ORE}(a)) \cite{Park2011}. In the presence of a chiral OAM texture, a shift of the occupation function by an external electric field results in a finite OAM, which is called the orbital Edelstein effect (OEE) (Fig.~\ref{fig:ORE}(b)) \cite{Go2017, Yoda2018, Salemi2019, Johansson2021}.

The ORE emerges even in the absence of  SOC, but the ORE complemented by SOC results in the ``spin'' Rashba effect \cite{Park2011, Sunko2017}, with the hierarchical relation similar to that of OHE and SHE. Thus, the ORE provides a coherent physical description of the electronic states with finite OAM and spin polarization. The ORE has been investigated in a wide range of systems, such as surfaces of metals \cite{Kim2012}, oxides \cite{Sunko2017}, topological insulators \cite{Park2012b}, and oxide interfaces \cite{Johansson2021}. Experimentally, the ORE has been verified by circular-dichroism in angle-resolved photoelectron spectroscopy \cite{Park2012a, Park2012b, Kim2012, Park2015, Sunko2017, Unzelmann2020} and momentum microscopy \cite{Tusche2015}. 

However, studies of transport effects in the presence of  ORE are limited and further research in this direction is required. For the OEE, Salemi \emph{et al.} investigated bulk collinear antiferromagnets Mn$_2$Au and CuMnAs \cite{Salemi2019}, and Johansson \emph{et al.} studied SrTiO$_3$ interfaces \cite{Johansson2021}. While these works consider only the atomic contribution, Niu \emph{et al.} investigated the OEE in mixed topological semimetals using Berry phase theory which consistently incorporates intra-atomic and inter-atomic contributions, observing a drastic variation in OEE for different angles of the magnetization in response to changes in the topology of the bands \cite{Niu2019}. For thin films, there can be an orbital current or diffusion of the OAM along the  direction  perpendicular to the surface/interface, which has been investigated for a surface-oxidized Cu \cite{Go2021a}.

\begin{figure}[b!]
\centering
\includegraphics[angle=0, width=0.25\textwidth]{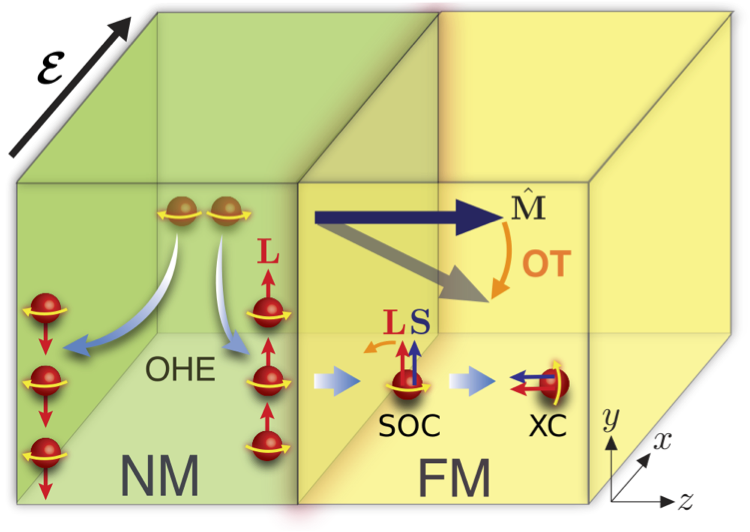}
\caption{
\label{fig:OT}
Mechanism of the OT in a NM/FM bilayer. The OHE in the NM results in the orbital injection (red arrow) into the FM, where it couples to spin (blue arrow) via SOC. Thus,  spin-orbit entangled states exert a torque on the magnetization. Adapted from Ref. \cite{Go2020a}.
}
\end{figure}

\section{Orbital torque}

Since the orbital current carries angular momentum as the spin current does, an injection of an orbital current into a FM can induce a torque on the magnetization. This type of torque induced by the orbital injection is named orbital torque (OT) \cite{Go2020a}. Phenomenologically, the OT is similar to the spin-transfer torque (ST) generated by spin injection \cite{Slonczewski1996, Berger1996}, but there are clear differences in microscopic mechanisms. Because  OAM cannot interact with the local magnetic moment directly, the OT requires  SOC in the FM. Thus, within the mechanism of the OT driven by the orbital current injected into the FM e.g. via the OHE in the NM, spin-orbit entangled states exert a torque on the magnetization (Fig.~\ref{fig:OT})  \cite{Go2020a}. In contrast, within the mechanism of the ST, the spin current alone is absorbed by the FM without a need for SOC. 

An important feature of the OT is that it is expected to depend sensitively on the chosen FM as the OT depends on the SOC in the FM \cite{Go2020a, Go2020b}, which is not expected for the ST \cite{Slonczewski1996, Berger1996}. In a magnetic heterostructure consisting of a NM layer and a FM layer, the NM generates both OHE and SHE, which result in the OT and ST, respectively. As a result, the sum of the OT and ST effectively acts on the magnetization. The relative sign of the OT and ST depends  not only on the spin-orbit correlation in the NM, which determines the relative sign between the OHE and SHE (Fig.~\ref{fig:OHE_SHE_VHE}(a)), but also that in the FM. The latter determines the ``orbital-to-spin'' conversion \cite{Go2020a, Go2020b}. For example, the spin-orbit correlation is expected to be positive for typical $3d$ FMs (Fe, Co, and Ni) but negative for Gd. Utilizing the OT instead of, or together with  ST may help to overcome material limitations  as the OHE has a much higher efficiency than the SHE, and even  small SOC of the FM may result in a sizeable OT. Thus, it is crucial to explore novel mechanisms of harnessing orbital currents, which have not been exploited so far, for the magnetization control in spintronics. 

\begin{figure}[b!]
\centering
\includegraphics[angle=0, width=0.49\textwidth]{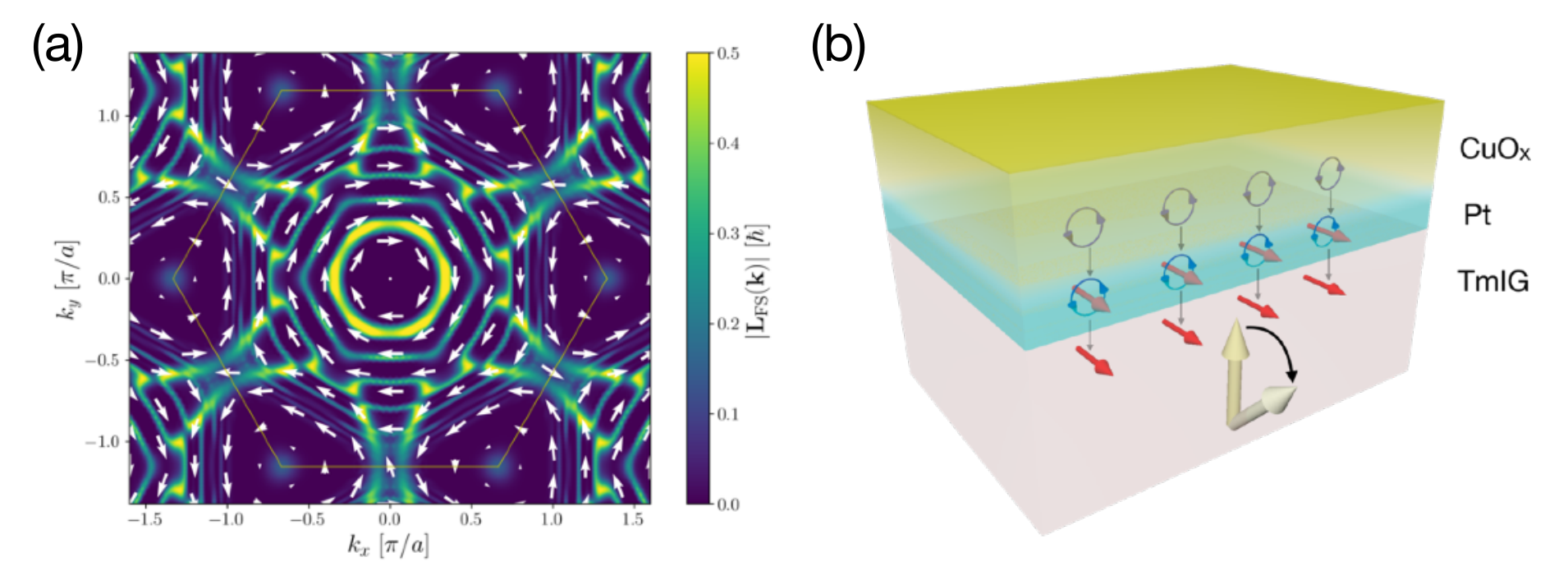}
\caption{
\label{fig:CuO}
(a) The $\mathbf{k}$-space OAM texture in a surface-oxidized Cu film at the Fermi surface. (b) The OT in a CuO$_x$/Pt/TmIG structure, where the OAM induced by the OEE (blue circular loops) strongly couples to the spin (red arrows) in an ultrathin Pt layer, which ultimately exerts a torque on the magnetic moment of TmIG (yellow arrows). Adapted from Refs. \cite{Go2021a, Ding2020}
}
\end{figure}

\subsection{Experimental evidence}

Remarkably, soon after the theoretical prediction, evidence of the OT has been presented for a few systems. One of them is a surface-oxidized Cu film. In 2016, An \emph{et al.} discovered that by naturally oxidizing the Cu layer in Cu/permalloy (Py) structures, the torque efficiency is strongly enhanced \cite{An2016}. This result was initially found to be quite surprising, as a material like Cu with negligible SOC is not expected to produce significant spin or its currents. After a few years a very prominent ORE was found theoretically in a surface-oxidized Cu film (Fig.~\ref{fig:CuO}(a)), which was shown  to generate not only a significant orbital accumulation but also a large orbital current flowing perpendicular to the interface \cite{Go2021a}. Experimentally, the orbital current mechanism was confirmed by Ding \emph{et al.} in CuO$_x$/Pt/TmIG, where the orbital current generated from  CuO$_x$ is efficiently converted into the spin current in an ultrathin ($\sim 1.5\ \mathrm{nm}$) Pt layer (Fig.~\ref{fig:CuO}(b)) \cite{Ding2020}. Meanwhile, Kim \emph{et al.} independently discovered that AlO$_x$/Cu/FM structures show a very large torque efficiency after a  partial oxidization by AlO$_x$ \cite{Kim2021}. An interesting feature found in this experiment is that the \emph{orbital transparency} depends significantly on the crystallinity of the Cu/FM interface, which was theoretically speculated to be very important for the OT mechanism \cite{Go2020a}. We note that Tazaki \emph{et al.} also found evidence of OT in CuO$_x$/FM structures, where the torque efficiency exhibits different sign depending on the FM choice \cite{Tazaki2020}.

Evidence for  OT was also found in NM/FM bilayers, where the orbital current is generated by the OHE in the NM bulk. Lee \emph{et al.} investigated various Cr-based heterostructures, where the variation in the sign and magnitude of the torque efficiency can be explained by the OT or the competition between the OT and ST \cite{Lee2021a} (note that the OHE and SHE have an opposite sign in Cr \cite{Jo2018}). Similar to Ref.~\cite{Ding2020}, it was found that the torque efficiency is significantly modified in Cr/Pt/CoFeB by the orbital-to-spin conversion in a Pt layer, when  compared to Cr/CoFeB. Meanwhile, Lee \emph{et al.} showed that the OT can be comparable to the ST even when the NM is a heavy element such as Ta, whose OHC is an order of magnitude larger than the SHC with opposite sign \cite{Lee2021b}. It has been shown that the torque efficiency changes sign in Ta/FM structures depending on the choice of the FM, which does not occur for Pt/FM structures since the OHC and SHC have same sign in Pt \cite{Lee2021b}. We note that the current-induced torque measured in Zr/CoFeB raised a possibility of the OHE from Zr because the SHE in Zr is far too small to explain the measured torque \cite{Zheng2020}. Meanwhile, a possible role of the ORE for current-induced torque in Pt/Co/SiO$_2$ was reported, where the inversion asymmetry leads to distortion of orbitals at the interface \cite{Chen2018}. 

An additional class of effects that relies on the transfer of angular momentum and related to OT are magnetoresistance effects such as the spin Hall magnetoresistance (SMR) \cite{Nakayama2013, Kim2016}, which results from the interaction of a spin current and the magnetization. Considering that an orbital current also interacts with the magnetization, a light metal or metal oxide where no significant SHE is expected can be used. This has been recently realized in a Py/CuO$_x$~\cite{Ding2021}. Here it was shown that the resistance of a the multilayer depends on the magnetization direction of Py with an angular dependence reminiscent of SMR. Given the negligible spin current generated from CuO$_x$, this could be ascribed to the orbital current induced by the ORE, which then interacts with the magnetization in Py via SOC. This effect was coined as orbital Rashba-Edelstein magnetoresistance and it is expected to occur in a wide range of systems which can host  orbital currents.

\section{Concluding remarks and outlook}

As the field of orbital transport is in its infancy, fundamental physical principles and novel transport effects await to be discovered. 
One of the most important tasks is to establish clear characterization tools in experiments. To date, no transport measurement scheme to detect the orbital currents, which does not involve spin currents, has been developed. While Xiao \emph{et al.} proposed a magnetoresistance measurement with a FM lead for the detection of the OHE \cite{Xiao2020}, it requires the SOC for the orbital-to-spin conversion, which is hard to disentangle from the signal by the SHE. One of the main problems here is that no  ``orbital polarizer'', which plays a role of a FM for the spin current, is known.

Another important phenomenon that needs to be addressed is the relaxation and dephasing of the orbital current. So far, there is no consensus on how far the orbital current can travel in different systems and setups and which microscopic processes lead to orbital relaxation. A recent theory suggests that the orbital current can propagate over much longer distances than the spin current in FMs, whose dephasing mechanism is completely different from that of the spin current \cite{Go2021b}. Interestingly,  recent magnetoresistance measurements in CuO$_x$/Py found that the dephasing length extracted by assuming the spin current model is much longer than the known value in Py \cite{Ding2021}. Considering that the orbital current is expected to be efficiently generated at the interface of CuO$_x$, the result suggests that the orbital current is absorbed more slowly than the spin current in FMs.

Considering the interaction of  electrons with other quasi-particles, one of the exciting directions to pursue lies with   effects due to interaction of  orbital currents with phonons and magnons. We remark that phononic analog of the OHE has been theoretically proposed recently \cite{Park2020}, and various mechanisms for inducing phonon OAM are being actively investigated \cite{Zhang2014, Juraschek2019, Hamada2018, Hamada2020}. In magnonics, it was suggested that the magnon Hall effect can be orbitally-polarized via spin excitations which carry spin chirality, with the magnon current ``dragging'' the electronic OAM \cite{Zhang2020}. Also, a concept of  magnonic OAM was proposed recently \cite{Neumann2020}. We remark that there exists an elementary excitation of orbital wave in an orbitally-ordered state \cite{Ishihara1997}, and such ``orbitons'' were measured in LaMnO$_3$ \cite{Saitoh2001}. We believe that studying orbital transport effects in transition metal oxides with orbital ordering may shed new light on the orbital physics and correlated phenomena \cite{Tokura2000, Khomskii2021}.

To summarize, the orbital current holds promises for new types of transport effects, which may possibly lead to new types of device applications that employ the orbital degree of freedom as an information carrier. The field is evolving rapidly in part due to an intensive  interaction with neighboring research areas such as phononics, spintronics and valleytronics, where the orbital current provides new twists and solutions to existing problems with spin and valley currents. We look forward to further explorations in this  exciting area of physics  in the near future.

\acknowledgments
This work was funded by the Deutsche Forschungsgemeinschaft (DFG, German Research Foundation) $-$ TRR173 $-$ 268565370 (projects A01, B02 and A11), TRR288 $-$ 422213477 (project B06). D.J. was supported by the Global Ph.D. Fellowship Program by National Research Foundation of Korea (Grant No. 2018H1A2A1060270). D.J. and H.-W.L acknowledge the financial support from the Samsung Science and Technology Foundation (Grant No. BA-1501-51).

\bibliographystyle{eplbib.bst}

\end{document}